# New Binary-Addition Tree Algorithm for the All-Multiterminal Binary-State Network Reliability Problem

Wei-Chang Yeh

Department of Industrial Engineering and Engineering Management

National Tsing Hua University

P.O. Box 24-60, Hsinchu, Taiwan 300, R.O.C.

yeh@ieee.org

*Abstract* — Various real-life applications, for example, Internet of Things, wireless sensor networks, smart grids, transportation networks, communication networks, social networks, and computer grid systems, are always modeled as network structures. The binary-state network composed of binary-state (e.g., functioning or failed) components (arcs and/or nodes) is one of the most popular network structures. The two-terminal network reliability is a success probability that the network is still functioning and can be calculated by verifying the connectivity between two specific nodes, and is an effective and popular technique for evaluating the performance of all types of networks. To obtain complete information for a making better decisions, a multi-terminal network reliability extends the two specific nodes to a specific node subset in which all nodes are connected. In this study, a new algorithm called the all-multiterminal BAT is proposed by revising the binary-addition-tree algorithm (BAT) and the layered-search algorithm (LSA) to calculate all multi-terminal reliabilities. The efficiency and effectiveness of the proposed all-multiterminal BAT are analyzed from the perspective of time complexity and explained via numerical experiments to solve the all-multiterminal network reliability problems.

## 1. INTRODUCTION

Owing to the universality, simplicity, and versatility of the network, real-world complex systems can be modelled in terms of network to discover practical properties and dynamical characteristics [1, 2]. Therefore, the network model has been widely researched so as to apply and implement it in areas of planning, designing, executing, managing, and controlling real-world complex systems [3, 4].

The binary-state network is a vital network model; each of its components is binary-state, either functioning or failed to communicate to send power [5], liquids [6], gases [6], data [7], topology design [8], signals [8, 9], resilience [10, 11], and multimedia [12]. Hence, the binary-state network has been applied to a variety of real applications, including communication [5, 8], transportation [6, 13], transformation [14], transmission [15], distribution [16], recovery [17], backup [18, 19], and allocation [20].

Network reliability is one of the most useful and important indices for measuring the performance of networks [2, 3, 21]. Network reliability is the success probability that the network is still functioning. Because whether the network is functioning is the most critical issue in network applications, reliability has been widely used to measure network performance in recent decades [21-24].

To calculate the network reliability, for example, the binary-state network reliability, is an NP-hard problem [2, 3, 4]. Various algorithms have been suggested to calculate the exact reliability [25-29] and estimate the approximate reliability of the binary-state network reliability [23, 30-32]. However, regardless of the former or the latter, all major previous studies are focused on the one-to-one reliability or two-terminal reliability. The two-terminal reliability [30, 32, 33] which involves finding the probability of two specific nodes, called the source node and the sink node, which are connected so as to communicate with each other.

Two-terminal binary-state reliability is fundamental to all types of network reliability [32-37]. The extension of two-terminal binary-state reliability is discussed below.

1. All two-terminal reliabilities (also called the all-pair reliability) calculate the two-terminal probability for each pair of nodes. Note that the number of specific nodes is limited to two in this category [37].

2. Many-to-many reliability, including one-to-many, one-to-all, many-to-one, and all-to-one, calculates the reliability between two node subsets. To calculate many-to-many reliability, each node in the first specific node subset must be connected to any node in the second specific node subset. However, there is no need to connect all the nodes within the first or second specific node subset. Note that the number of specific nodes is no longer limited to two in this category [33].

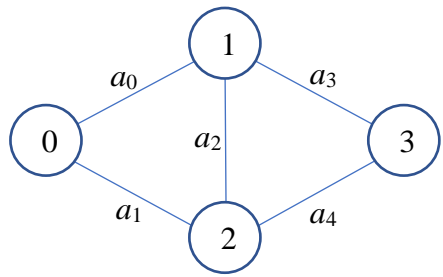

**Figure 1.** Example undirect graph.

For example, the reliability between nodes 0 and 3 in the binary-state network shown in Fig. 1 is called the two-terminal reliability; all reliabilities between nodes 0 and 1, nodes 1 and 2, nodes 0 and 3, nodes 1 and 2, nodes 1 and 3, and nodes 2 and 3 are called the all-pair reliability; the reliability between node 0 and node subset {1, 3} is called the one-to-many reliability; the reliability between node {0, 2} and node subset {1, 3} is called the many-to-many reliability.

Regardless of the number of specific nodes, all binary-state reliability problems mainly focus only on the node subset included source node and another node subset included the sink side, and each subset can have more than one node [32, 33, 37]. Hence, the traditional binary-state reliability calculates the connectivity between two node subsets such as $V^*$ and $V^{\#}$ such that $V^* \cap V^{\#} = \varnothing$, $V^* \cup V^{\#} \subseteq V$ (the node set), and there is a connection between nodes $v^*$ and $v^{\#}$ for all $v^* \in V^*$ and $v^{\#} \in V^{\#}$ if the network is functioning. Note that all nodes in $V^*$ or $V^{\#}$ can be disconnected.

Thus, the traditional generalized binary-state reliability is called the two-subset reliability here because these two node subsets are located on two distinctive and disjoint (node) subsets.

However, the two-subset reliability is not sufficient to measure the network performance and the roles of components in making a correct decision. For example, a decision-maker is always needed to find which of the four locations is the most reliable for building a supply chain in the transportation system [13]; which five bases have the best reliable connection after considering network resiliency in telecom systems [10, 11]; which three nodes have less reliable communication in the wireless sensor network [8, 9, 10, 16], which six grids need to enhance their connection to have a better reliable smart grid [5, 20]; and which five points have the weakest signal transmission in the Internet of Things systems [15, 38].

Hence, a novel binary-state reliability problem called the all-multiterminal reliability problem is proposed in this study to overcome the insufficiency of the traditional two-subset reliability.

In the all-multiterminal reliability problem, we need to find the connection between all nodes in a specific node subset, that is, all nodes in the specific node subset $V^* \subseteq V$ are connected. Note that there are two major differences between the proposed all-multiterminal reliability problem and two-subset reliability problem: the former has only one specific node subset and all nodes in such node subset must be connected to each other; the latter has two specific node subsets and all nodes within these two node subsets can be disconnected to each other, and each node is connected to at least one node in another specific node subset. Thus, the all-multiterminal network reliability cannot be solved by applying an algorithm that calculates the two-subset binary-state network reliability. To solve the proposed novel problem, a new algorithm called the all-multiterminal BAT based on the binary-addition-tree algorithm (BAT) [39-47] is developed.

Thus, the major contributions of this study are the proposal of a novel all-multiterminal reliability problem for a binary-state network together with a new BAT-based method to calculate all-multiterminal reliability. The remainder of this paper is organized as follows. Section 2 introduces acronyms, notations, nomenclatures, and assumptions used in this study. A review of the traditional BAT, including the arc-based and node-based BATs and PLSA [47, 51], are discussed in Section 3. The proposed TLSA used to verify the connections among all nodes in a specific node subset is provided in Section 4. The proposed proper-subset BAT for finding all non-empty node subsets in the node subset is presented in Section 5. The proposed all-multiterminal BAT integrated with the traditional arc-based BAT, TLSA, and proper-subset BAT are discussed in Section 6, together with an example to demonstrate its performance. Finally, in Section 6, we present our conclusions.

## 2. ACRONYMS, NOTATIONS, NOMENCLATURES, AND ASSUMPTIONS

The required acronyms, notations, assumptions, and nomenclatures are presented in this section.

### 2.1 Acronyms

BAT: binary-addition tree algorithm [47]

LSA: layered-search algorithm [51]

PLSA: path-based LSA [47]

TLSA: tree-based LSA

## 2.2 Notations

$/\bullet/$: number of elements in $\bullet$

$V$: node set $V = \{0, 1, 2, \ldots, (n-1)\}$

$E$: arc set $E = \{a_0, a_1, \ldots, a_{m-1}\}$

$a_i$: arc $a_i \in E$ for $i = 0, 1, \ldots, (m-1)$

$n$: number of nodes $n = |V|$

$m$: number of arcs $m = |E|$

**D**: state distribution involving states and the corresponding probability of each arc. For example, **Table 1** is a binary-state distribution for Fig. 1.

**Table 2.** Binary-state distribution **D** in Fig. 1

| $i$ | $\Pr(a_i)$ |
|-----|------------|
| 0 | 0.9 |
| 1 | 0.8 |
| 2 | 0.7 |
| 3 | 0.7 |
| 4 | 0.8 |

$G(V, E)$: An undirect graph with node set $V$ and arc set $E$. For example, Fig. 1 is an undirect graph with $V = \{0, 1, 2, 3\}$ and $E = \{a_0, a_1, a_2, a_3, a_4\}$.

$G(V, E, \mathbf{D})$: A binary-state network with **D** and $G(V, E)$. For example, Fig. 1 is a binary-state network after **D** is given in Table 1.

$X$: ($m$-tuple) binary-state vector obtained from the forward binary-state BAT

$\underline{X}$: ($m$-tuple) binary-state vector obtained from the backward binary-state BAT

$X(a_i)$: value of the coordinate $i$ for $i = 0, 1, \ldots, (m-1)$, e.g., $X(a_0) = X(a_1) = X(a_2) = 1$ and $X(a_3) = X(a_4) = 0$ if $X = (1, 1, 1, 0, 0)$.

$G(X)$: subgraph $G(X) = G(V, E^*)$, where $E^* = \{a \in E \mid \text{for all } a \text{ with } X(a) = 1\}$. For example, $G(X_7)$ is depicted in Fig. 2, where $X_7 = (1, 1, 1, 0, 0)$ and $G(V, E)$ is shown in Fig. 1.

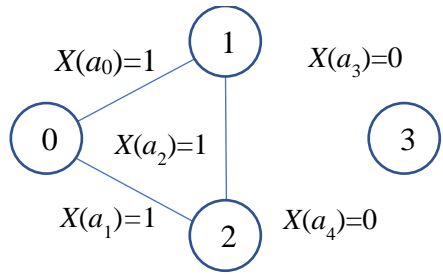

**Figure 2.** $X_7 = (1, 1, 1, 0, 0)$ and $G(X_7)$ in Fig. 1.

$W_f(X)$:    $W_f(X) = \sum_{i=0}^{m-1} 2^i X(a_i)$ [51]

$W_b(X)$:    $W_b(X) = \sum_{i=0}^{m-1} 2^{m-i-1} X(a_i)$ [51]

$\Pr(\bullet)$:   probability to have $\bullet$

$\Pr_p(\bullet)$:   probability to have $\bullet$ with $\Pr(a) = p$ for all $a \in E$

$\Pr(X)$:   $\Pr(X) = \Pr(X(a_0)) \cdot \Pr(X(a_1)) \cdot \ldots \cdot \Pr(X(a_{m-1}))$ for each binary-state vector $X$

$\Pr_p(X)$:   $\Pr_p(X) = \Pr_p(X(a_0)) \cdot \Pr_p(X(a_1)) \cdot \ldots \cdot \Pr_p(X(a_{m-1}))$ for each binary-state vector $X$

$R(\bullet)$:   reliability of $G(\bullet, E^*, \mathbf{D})$, where $\bullet$ is a node subset of $V$ and $E^* = \{\ a \in E \mid$ if the endpoints of arc $a$ are both in $\bullet\}$.

$A << B$:  vector $A$ is obtained earlier than vector $B$ in the procedures of BAT-based algorithms

## 2.3 Nomenclatures

Two-terminal reliability:   The success probability that two specific nodes are connected in a network [30, 32, 33].

All-pair reliability:   All two-terminal reliabilities [37].

Proper subset:   A non-empty subset.

Multi-terminal reliability:   The success probability that all nodes in a given node subset are connected in a network.

All-multiterminal reliability: All multi-terminal reliabilities.

## 2.4 Assumptions

1. Each node is completely reliable and connected to all the nodes in $G(V, E)$.

2. Each arc $a \in E$ is binary-state, that is, it has two states either functioning or failed, and $\Pr(a)$ is statistically independent according to $\mathbf{D}$ in $G(V, E)$.

3. $G(V, E)$ has no parallel arcs or loops.

## 3. REVIEW OF BAT AND PLSA

The proposed new all-multiterminal BAT is based on the (traditional) arc-based BAT [47] to find state vectors. The proposed proper-subset BAT is based on node-based BAT [41] to find all proper node subsets in a connected node subset, and the proposed TLSA is based on the PLSA [47, 51] to find connected disjoint node subsets. In Section 3, we briefly describe the arc-based BAT, node-based BAT, and PLSA.

### 3.1 Arc-Based BAT and Node-Based BAT

Both the arc-based BAT and node-based methods were recently proposed by Yeh in [47] and [41] to generate all arc-based and node-based binary-state vectors, respectively. Each coordinate, say $i$, of an arc-based binary-state vector is either 0 or 1, representing whether arc $i$ fails or works for $i = 0$, 1, ..., $(m-1)$. Similarly, each coordinate, say $j$, of a node-based binary-state vector is either 0 or 1, representing whether node $j$ belongs to the related node subset for $i = 0$, 1, ..., $(n^*-1)$, where $n^*$ is the number of nodes in the node subset we are concerned.

All BATs, regardless of whether they are arc-based or node-based, simply list all state vectors by a procedure analogous to adding one to a binary code, that is, all BATs are basically established on the binary addition operator to find all binary-state vectors [39-47].

There are two different kinds of BATs in terms of adding one to a binary-state vector: the backward BAT and the forward BAT. The former adds one to the last coordinate and gradually moves to the first coordinate of the binary-state vectors. Conversely, the latter adds one to the first coordinate and progressively moves to the last coordinate of the binary-state vectors. The forward BAT is adopted in this study, and its source code can be downloaded from [47].

The forward arc-based BAT [47] is provided below.

**Algorithm: the forward arc-based BAT**

**Input:** $G(V, E)$.

**Output:** All arc-based binary-state vectors.

**STEP F0.** Let $i$ and $X$ be zero and zero vector, respectively.

**STEP F1.** If $X(a_i) = 0$, go to STEP F3.

**STEP F2.** Let $X(a_i) = 0$, $i = i + 1$, and go to STEP F1.

**STEP F3.** If $i = m$, halt. Otherwise, let $X(a_i) = 1$, $i = 0$, we have a new binary-state vector $X$, and go to STEP F1.

Note that "$i = m$" is change to "$i = n^*$" in STEP F3 in the node-based BAT.

In STEP F0 of the forward arc-based BAT pseudocode, the current coordinate and the current binary-state vector are initialized. For instance, the first coordinate $i = 0$ and the first 5-tuple vector $X = (0, 0, 0, 0, 0)$ in Fig. 1.

The loop from STEPs F1 to F3 generates all binary-state vectors. STEP F3 is the stopping criterion, and it stops the entire procedure if $i$ reaches the coordinate $m$, for example, the next vector after $X = (1, 1, 1, 1, 1)$ is $(0, 0, 0, 0, 0, 1)$, which is a $(m+1)$-tuple coordinate and halt in Fig. 1. Otherwise, the value of the current coordinate is changed from zero to one, a new vector $X$ is generated, and go to STEP F1 after resetting $i = 0$, for example, $X = (0, 0, 1, 1, 0)$ is changed to $(1, 0, 1, 1, 0)$, $i$ is set to 0, and go to STEP F1.

STEP F2 changes the value of the current coordinate from one to zero and move to the next coordinate. For example, in Fig. 1, $X = (1, 1, 1, 1, 0)$ is changed to $(0, 1, 1, 1, 0)$, $(0, 0, 1, 1, 0)$, $(0, 0, 0, 1, 0)$, $(0, 0, 0, 0, 1)$ in the loop, and return to STEP F1.

Both forward and backward BATs have time complexity $(2^m + 1)$ [37]. Because BAT is easy to code, flexible to make-to-fit, and more efficient than other search methods such as DFS and BFS. Hence, various BATs have been proposed for practical numerical problems immediately after the emergence of BAT.

One common shared with all various binary-state BATs is listed below [49], where $X$ and $Y$ are two vectors obtained from BAT:

for all forward BAT

$$W_f(X) \; = \sum_{i=0}^{m-1} 2^i X(a_i) < W_f(Y) \; = \sum_{i=0}^{m-1} 2^i Y(a_i) \text{ if and only if } X \ll Y; \qquad (1)$$

for all backward BAT:

$$W_b(X) = \sum_{i=0}^{m-1} 2^{m-i+1} X(a_i) < W_b(Y) = \sum_{i=0}^{m-1} 2^{m-i+1} Y(a_i) \text{ if and only if } X \ll Y. \quad (2)$$

Note that $W_f(X) = W_b(\underline{X})$ and $X(a_i) = \underline{X}(a_{m-i+1})$ for $i = 0, 1, \ldots, (m-1)$.

Table 3 records $X_i$, $B(X_i)$, $W_f(X_i)$, $\mathrm{Pr}_{0.9}(X_i)$, $\underline{X}_i$, $B(\underline{X}_i)$, $W_b(\underline{X}_i)$, and $\mathrm{Pr}_{0.9}(\underline{X}_i)$, in Fig. 1 for $i = 0, 1, \ldots,$ 31, where $X_i$ and $\underline{X}_i$ are the $i$th binary-state vectors obtained from the forward binary-state BAT and forward binary-state BAT, respectively [49]. For example, $X_7 = (1, 1, 1, 0, 0)$ is obtained in the 7th iteration from the forward BAT, $B(X_7) = 11100$, $W_f(X_7) = W_b(\underline{X}_7) = 7$, $\mathrm{Pr}_{0.9}(X_7) = \mathrm{Pr}_{0.9}(\underline{X}_7) = 0.9^3 \times 0.1^2 = 0.00729$, the 7th vector obtained from the backward BAT is $\underline{X}_7 = (0, 0, 1, 1, 1)$, and $B(\underline{X}_7) = 00111$.

**Table 3.** All vectors and related information obtained from the binary-state BAT.

| $i$ | $X_i$ | $B(X_i)$ | $W_f(X_i) = W_b(\underline{X}_i)$ | $\mathrm{Pr}_{0.9}(X_i) = \mathrm{Pr}_{0.9}(\underline{X}_i)$ | $\underline{X}_i$ | $B(\underline{X}_i)$ |
|---|---|---|---|---|---|---|
| 0 | (0, 0, 0, 0, 0) | 00000 | 0 | 0.000010 | (0, 0, 0, 0, 0) | 00000 |
| 1 | (1, 0, 0, 0, 0) | 10000 | 1 | 0.000090 | (0, 0, 0, 0, 1) | 00001 |
| 2 | (0, 1, 0, 0, 0) | 01000 | 2 | 0.000090 | (0, 0, 0, 1, 0) | 00010 |
| 3 | (1, 1, 0, 0, 0) | 11000 | 3 | 0.000810 | (0, 0, 0, 1, 1) | 00011 |
| 4 | (0, 0, 1, 0, 0) | 00100 | 4 | 0.000090 | (0, 0, 1, 0, 0) | 00100 |
| 5 | (1, 0, 1, 0, 0) | 10100 | 5 | 0.000810 | (0, 0, 1, 0, 1) | 00101 |
| 6 | (0, 1, 1, 0, 0) | 01100 | 6 | 0.000810 | (0, 0, 1, 1, 0) | 00110 |
| 7 | (1, 1, 1, 0, 0) | 11100 | 7 | 0.007290 | (0, 0, 1, 1, 1) | 00111 |
| 8 | (0, 0, 0, 1, 0) | 00010 | 8 | 0.000090 | (0, 1, 0, 0, 0) | 01000 |
| 9 | (1, 0, 0, 1, 0) | 10010 | 9 | 0.000810 | (0, 1, 0, 0, 1) | 01001 |
| 10 | (0, 1, 0, 1, 0) | 01010 | 10 | 0.000810 | (0, 1, 0, 1, 0) | 01010 |
| 11 | (1, 1, 0, 1, 0) | 11010 | 11 | 0.007290 | (0, 1, 0, 1, 1) | 01011 |
| 12 | (0, 0, 1, 1, 0) | 00110 | 12 | 0.000810 | (0, 1, 1, 0, 0) | 01100 |
| 13 | (1, 0, 1, 1, 0) | 10110 | 13 | 0.007290 | (0, 1, 1, 0, 1) | 01101 |
| 14 | (0, 1, 1, 1, 0) | 01110 | 14 | 0.007290 | (0, 1, 1, 1, 0) | 01110 |
| 15 | (1, 1, 1, 1, 0) | 11110 | 15 | 0.065610 | (0, 1, 1, 1, 1) | 01111 |
| 16 | (0, 0, 0, 0, 1) | 00001 | 16 | 0.000090 | (1, 0, 0, 0, 0) | 10000 |
| 17 | (1, 0, 0, 0, 1) | 10001 | 17 | 0.000810 | (1, 0, 0, 0, 1) | 10001 |
| 18 | (0, 1, 0, 0, 1) | 01001 | 18 | 0.000810 | (1, 0, 0, 1, 0) | 10010 |
| 19 | (1, 1, 0, 0, 1) | 11001 | 19 | 0.007290 | (1, 0, 0, 1, 1) | 10011 |
| 20 | (0, 0, 1, 0, 1) | 00101 | 20 | 0.000810 | (1, 0, 1, 0, 0) | 10100 |
| 21 | (1, 0, 1, 0, 1) | 10101 | 21 | 0.007290 | (1, 0, 1, 0, 1) | 10101 |
| 22 | (0, 1, 1, 0, 1) | 01101 | 22 | 0.007290 | (1, 0, 1, 1, 0) | 10110 |
| 23 | (1, 1, 1, 0, 1) | 11101 | 23 | 0.065610 | (1, 0, 1, 1, 1) | 10111 |
| 24 | (0, 0, 0, 1, 1) | 00011 | 24 | 0.000810 | (1, 1, 0, 0, 0) | 11000 |
| 25 | (1, 0, 0, 1, 1) | 10011 | 25 | 0.007290 | (1, 1, 0, 0, 1) | 11001 |
| 26 | (0, 1, 0, 1, 1) | 01011 | 26 | 0.007290 | (1, 1, 0, 1, 0) | 11010 |
| 27 | (1, 1, 0, 1, 1) | 11011 | 27 | 0.065610 | (1, 1, 0, 1, 1) | 11011 |
| 28 | (0, 0, 1, 1, 1) | 00111 | 28 | 0.007290 | (1, 1, 1, 0, 0) | 11100 |
| 29 | (1, 0, 1, 1, 1) | 10111 | 29 | 0.065610 | (1, 1, 1, 0, 1) | 11101 |
| 30 | (0, 1, 1, 1, 1) | 01111 | 30 | 0.065610 | (1, 1, 1, 1, 0) | 11110 |
| 31 | (1, 1, 1, 1, 1) | 11111 | 31 | 0.590490 | (1, 1, 1, 1, 1) | 11111 |
| | SUM | | | 1.000000 | | |

**3.2 PLSA**

The PLSA [47] is extended from the layered-search algorithm (LSA), which was originally proposed in [51] to find all $d$-MPs in acyclic networks. The basic idea of the LSA is to find disjoint layers such that the first layer contains the source node only, the $i$th layer includes all nodes connected to all nodes in the $(i+1)$th layer, and to determine whether there are $d$ paths from the source node to the sink node [51].

The above concept in LSA is simplified and implemented in the PLSA to verify the connectivity of two specific nodes in two-terminal network reliability problems [47, 51]. If the last layer contains a sink node, the source node is connected to the sink node. Otherwise, the last layer is empty, and the sink node is not included in the last layer, that is, the source and sink nodes are disconnected.

Because two-terminal network reliability is the probability that two specific nodes are connected, there is always a need to verify the connectivity of the source and sink nodes [30, 32, 33]. In addition, the PLSA is very efficient, simple, and easy to code, and it is continuously implemented to verify the connectivity of two specific nodes in each sub-network represented by the state vector in most of BAT-based algorithms [39-47].

The pseudocode of the PLSA is provided below to establish the connectivity of the source node and the sink node in $G(X)$, where $X$ can be any binary-state or multi-state vector obtained from the BAT in a graph that can be binary-state and multi-state, respectively [48]:

**Algorithm: PLSA**

**Input:**     A vector $X$ in $G(V, E)$, two specific nodes $s$ and $t$.

**Output:**   The connectivity of nodes $s$ and $t$ in $G(X)$.

**STEP P0.**  Let the layer index $i = 0$ and $L_0 = \{s\}$.

**STEP P1.**  Let $L_i = \{ v \notin (L_{i-1} - \bigcup_{k=0}^{i-2} L_k) \mid e_{l,v} \in G(X)$ and $l \in (L_{i-1} - \bigcup_{k=0}^{i-2} L_k) \} \subseteq G(X)$.

**STEP P2.**  If $t \in L_i$, halt and node $s$ and node $t$ are connected in $G(X)$.

**STEP P3.**  If $L_i = \varnothing$, halt and node $s$ and node $t$ are disconnected in $G(X)$. Otherwise, let $i = i + 1$ and go to STEP P1.

The time complexity of the PLSA is $O(n)$ because there are $n$ layers in STEP P1 of the above pseudocode in the worst case [47]. For example, the PLSA is implemented to verify the connectivity between nodes 0 and 4 in Fig. 3, and the procedure is shown in Table 4.

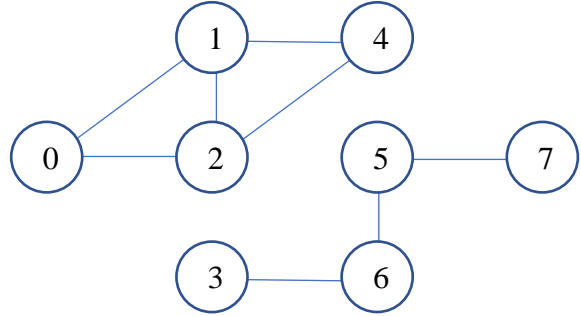

**Figure 3.** PLSA Example network.

**Table 4.** PLSA example procedure.

| $i$ | $L_i$ | Remark |
|---|---|---|
| 0 | {0} | |
| 1 | {1, 2} | |
| 2 | {4} | |
| 3 | $\varnothing$ | Nodes 0 and 4 are connected |

## 4 PROPOSED TLSA

Inherits the efficiency and simplicity of the LSA [47, 51], the PLSA is extended to the proposed TLSA to verify the connectivity of a node subset with more than two nodes by finding all disjoint trees in $G(X)$ for the state vector $X$. The details of the proposed TLSA are explained in this section.

### 4.1 Repeated PLSA to Verify the Connectivity of Node Subsets

Let $V^*$ be the node subset that we need to verify the connectivity among all nodes in $V^*$. We can implement the PLSA to each pair of nodes to verify their connectivity because all nodes are connected if all pairs of nodes are connected. Otherwise, all nodes are not connected.

Thus, we repeated PLSA, which re-implement the PLSA until all pairs of nodes are tested. The pseudocode for the repeated PLSA is as follows:

**Algorithm: Repeated PLSA**

**Input:**     $G(V, E)$, and $X$.

**Output:**    The connectivity of nodes in $G(X)$.

**STEP R0.** Let $i = 1$.

**STEP R1.** Let $j = i + 1$.

**STEP R2.** Implement PLSA to verify whether the $i$th node and the $j$th node are connected in $G(X)$.

**STEP R3.** If the $i$th node and the $j$th node are disconnected in $G(X)$, let Conn$(i, j) = 0$; else Conn$(i, j) = 1$.

**STEP R4.** If $j < n$, let $j = j + 1$ and go to STEP R2.

**STEP R5.** If $i < (n-1)$, let $i = i + 1$ and go to STEP R1. Otherwise, halt.

**STEP R6.** All nodes in $V^* \subseteq V$ are connected if Conn$(i, j) = 1$ for all nodes $i, j \in V$.

The above algorithm takes $O(n \times (n-1) \times n) \approx O(n^3)$ to verify the connectivity among nodes, where $O(n)$ is the time complexity of PLSA, and $O(n \times (n-1)) = O(n^2)$ is the number of combinations of distinctive pairs of nodes.

If we can reduce the time complexity in verifying the connectivity among nodes, we can speed up the BAT implemented in calculating the network reliability. To obtain a better algorithm to replace the repeated PLSA discussed above, a novel search method called tree-based LSA (TLSA) is proposed with time complexity $O(|V^*|)$, which is much better than that of the PLSA to determine the connectivity among nodes in $V^*$.

## 4.2 Proposed TLSA

The basic concept of the proposed TLSA is to implement the PLSA more efficiently than the repeated PLSA discussed in Section 4.1. In the proposed TLSA, PLSA finds distinctive connected subsets each time, and there are at most $n$ times to run the PLSA if all nodes are disconnected [48]. For example, PLSA starts from node $i$ and finds all nodes including node $i$ in $V_i$ that are connected to each other. Next time, PLSA starts from any node, that is, node $j \notin V_i$.

Continuing the above process, we can find all disjoint node subsets, all nodes in each node subset are connected to each other, and no two nodes are connected if these two nodes are in two different node subsets. Hence, nodes are searched at most once in PLSA, that is, the time complexity of the proposed TLSA is $O(n)$ because there are $n$ nodes only.

Hence, the time complexity is reduced from $O(n^3)$ in the repeated PLSA to $O(n)$ in the proposed TLSA, that is, TLSA is more efficient than repeated PLSA.

### 4.3 TLSA Pseudocode

The following is the proposed TLSA pseudocode:

**Algorithm: TLSA**

**Input:**        $G(V, E)$ and $X$.

**Output:**    All connected node subsets in $G(X)$.

**STEP T0.**  Let $j = k = 0$, and $V^* = V$.

**STEP T1.**  Implement the PLSA shown in Section 3.2 by letting the first layer $L_{k,0} = \{j\}$, $L_{k,h} = \{v \notin L_{k,h-1} \mid$ there is one arc in $E$ connected one node in $L_{k,h-1}$ and $v\} \subseteq V$, and assume that $L_{k,i} = \varnothing$ is the last layer.

**STEP T2.**  Let $V_k = L_{k,0} \cup L_{k,1} \cup \ldots \cup L_{k,(i-1)}$ and $V^* = V^* - V_k$.

**STEP T3.**  If $V^* = \varnothing$, halt.

**STEP T4.**  Let a new node $j \in V^*$, $k = k + 1$, and go to STEP T1.

STEP T0 starts from node 0, similar to the traditional PLSA. STEP T1 implements the traditional PLSA to find all layers connected from node $j$. STEP T2 unions all nodes in each layer to be $V_k$ and updates $V^*$ by subtracting $V_k$. STEP T3 checks whether $V^*$ is empty, which is also the stopping criterion of the proposed TLSA. STEP T4 repeats the above process by selecting a new node, that is, node $j$, in $V^*$. As discussed in Section 4.2, each node appears at most once in at most one layer, that is, the time complexity of the proposed TLSA is $O(n)$.

### 4.4 Example

The above TLSA pseudocode is demonstrated to explain how to be implemented to verify the connectivity of Fig. 4, and the related process is shown in Tables 4 and 5.

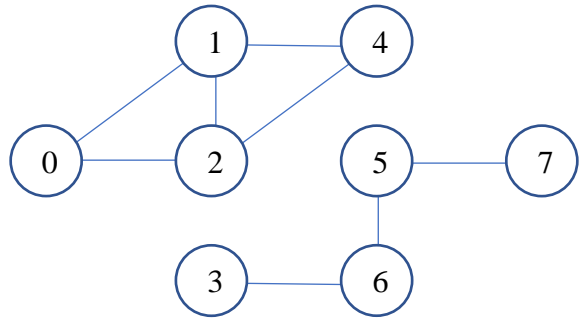

**Figure 4.** PLSA Example network.

**STEP T0.** Let $j = k = 0$ and $V^* = V = \{0, 1, 2, 3, 4, 5, 6, 7\}$.

**STEP T1.** Implementing the PLSA, as shown in Table 4 but adding a new subscript 0 in the first position to indicate that this is obtained from node 0 in the first PLSA. Then, we have $L_{0,0} = \{0\}$, $L_{0,1} = \{1, 2\}$, and $L_{0,2} = \{4\}$.

**STEP T2.** Let $V_0 = L_{0,0} \cup L_{0,1} \cup L_{0,2} = \{0, 1, 2, 4\}$, i.e., $\{0, 1, 2, 4\}$ is a connected subset, and $V^* = V^* - V_k = \{3, 5, 6, 7\}$.

**STEP T3.** Because $V^* \neq \varnothing$, go to STEP T4.

**STEP T4.** Let $k = k + 1 = 1$ and find any node, say node 3, in $V^* = \{3, 5, 6, 7\}$ and go to STEP T1.

**STEP T1.** After implementing the PLSA, as shown in Table 5, $L_{1,0} = \{3\}$, $L_{1,1} = \{6\}$, $L_{1,2} = \{5\}$, and $L_{1,3} = \{7\}$.

**STEP T2.** Let $V_k = L_{1,0} \cup L_{1,1} \cup L_{1,2} \cup L_{1,3} = \{3, 5, 6, 7\}$ which is a connected node subset and $V^* = V^* - V_k = \varnothing$.

**STEP T3.** Because $V^* = \varnothing$, halt.

**Table 5.** The procedure of the second PLSA.

| $i$ | $L_{1,i}$ | Remark |
|---|---|---|
| 0 | $\{3\}$ | |
| 1 | $\{6\}$ | |
| 2 | $\{5\}$ | |
| 3 | $\{7\}$ | |
| 4 | $\varnothing$ | $\{3, 5, 6, 7\}$ is connected. |

## 5. PROPOSED PROPER-SUBSET BAT

A proper subset is a subset that is not empty. Each proper subset of a connected node subset was also connected. Hence, after finding all connected disjoint subsets, we need to have their proper subsets to count into the multi-terminal reliability. For example, in Fig. 4, $\{3, 5, 6, 7\}$ is connected and all of its proper subsets, {3, 5}, {3, 6} {3, 7}, {5, 6}, {5, 7}, {6, 7}, {3, 5, 6}, {3, 5, 7}, {3, 6, 7}, {5, 6, 7}, and {3, 5, 6, 7}, are also connected. Each proper subset, for example, {3, 5}, needs to be considered when calculating $R_{\{3, 5\}}$.

Because

$$R_{V^*} = \sum_X \Pr(X) \text{ for all state vectors } X \text{ with node subset } V^* \text{ is connected in } G(X). \qquad (3)$$

We need to verify whether the non-empty node subset $V^*$ is connected in $G(X)$ for all $X$. The proposed TLSA can identify all the connected distinctive node subsets. After finding each connected node subset, we must have all the proper subsets of such node subsets.

To achieve this goal, a proper-subset BAT is proposed to find each proper subset of a connected node subset. Let $Y_{i,j}$ be the $j$th vector obtained in the connected node subset $V^*$ such that $k$ is included in the proper subset if $Y_{i,j}(k) = 1$; otherwise, $Y_{i,j}(k) = 0$. The pseudocode of the proposed proper-subset BAT is provided below.

**Algorithm: The Proper-subset BAT**

**Input:**     $G(V, E, \mathbf{D})$, $X$, and node subset $V^*$ is connected in $G(X)$.

**Output:**   All proper subset of $V^*$.

**STEP S0.**   Implement the node-based BAT to obtain each $|V^*|$-tuple vector, say $Y_{i,j}$ for all $j = 1, 2, \ldots, n_i$, where $n_i$ is the number of obtained $|V^*|$-tuple vectors.

**STEP S1.**   Let $j = 1$.

**STEP S2.**   Let $k = 1$ and $|Y_{i,j}| = 0$.

**STEP S3.**   Let $h$ be node of the $k$th coordinate in $Y_{i,j}$.

**STEP S4.**   If $Y_{i,j}(h) = 1$, let $|Y_{i,k}| = |Y_{i,k}| + 2^h$.

**STEP S5.**   If $k < |V^*|$, let $k = k + 1$ and go to STEP S3.

**STEP S6.**   Let $R_{|Y_{i,j}|} = R_{|Y_{i,j}|} + \Pr(X)$.

**STEP S7.**   If $j < n_i$, let $j = j + 1$ and go to STEP S2. Otherwise, halt.

Note that state vector $Y_{i,j}$ is a vectorized proper subset obtained from the proposed proper-subset BAT of a connected node subset $V_i$ in $G(X)$, where node subset $V_i$ is verified by the proposed TLSA, and state vector $X$ is obtained from the traditional BAT [47].

STEP S0 finds all vectorized subsets of $V^*$. The loop from STEPs S2 to S7 is the major contribution of the proper-subset BAT, and it finds the corresponding label of the related proper subset of $V^*$. This loop calculates the label, that is, $|Y_{i,j}|$ from STEPs S2 to S5, of the related proper subset of $V^*$ obtained in STEP S0. After obtaining the label, we can add $\Pr(X)$ to $R_{|X_{i,j}|}$ easily because it requires more time to locate the node subset with respect to $\{ k \mid X_{i,j}(k) = 1 \text{ for all } k \}$. Hence, without this loop, we need to compare the obtained proper subset of $V^*$ to each proper subset of $V$ to find the correct subset to add $\Pr(X)$.

The time complexity of the proposed proper-subset BAT is only $O(2^a)$ based on [37] in STEP S0, where $a = |V^*|$ and it takes $O(a)$ to calculate $|V_{i,j}|$ for an $a$-tuple vector. Hence, the time complexity is $O(a2^a)$ and $O(n2^n)$ for each $V^*$ and $V$.

For example, $V_0 = \{0, 1, 2, 4\}$ and $V_1 = \{3, 5, 6, 7\}$ are two connected disjoint node subsets obtained from the TLSA in Section 4.4. Each vectorized proper subset and its label are listed in Table 6, based on STEP S0 and STEPs S2 to S5.

**Table 6.** All vectorized proper subsets obtained from the proposed proper-subset BAT.

| $i$ | $Y_{0,i} = (y_0, y_1, y_2, y_4)$ | $|Y_{0,i}|$ | $Y_{1,i} = (y_3, y_5, y_6, y_7)$ | $|Y_{1,i}|$ |
|---|---|---|---|---|
| 1 | $(1, 0, 0, 0)$ | 1 | $(1, 0, 0, 0)$ | 8 |
| 2 | $(0, 1, 0, 0)$ | 2 | $(0, 1, 0, 0)$ | 32 |
| 3 | $(1, 1, 0, 0)$ | 3 | $(1, 1, 0, 0)$ | 40 |
| 4 | $(0, 0, 1, 0)$ | 4 | $(0, 0, 1, 0)$ | 64 |
| 5 | $(1, 0, 1, 0)$ | 5 | $(1, 0, 1, 0)$ | 72 |
| 6 | $(0, 1, 1, 0)$ | 6 | $(0, 1, 1, 0)$ | 96 |
| 7 | $(1, 1, 1, 0)$ | 7 | $(1, 1, 1, 0)$ | 104 |
| 8 | $(0, 0, 0, 1)$ | 16 | $(0, 0, 0, 1)$ | 128 |
| 9 | $(1, 0, 0, 1)$ | 17 | $(1, 0, 0, 1)$ | 136 |
| 10 | $(0, 1, 0, 1)$ | 18 | $(0, 1, 0, 1)$ | 160 |
| 11 | $(1, 1, 0, 1)$ | 19 | $(1, 1, 0, 1)$ | 168 |
| 12 | $(0, 0, 1, 1)$ | 20 | $(0, 0, 1, 1)$ | 192 |
| 13 | $(1, 0, 1, 1)$ | 21 | $(1, 0, 1, 1)$ | 200 |
| 14 | $(0, 1, 1, 1)$ | 22 | $(0, 1, 1, 1)$ | 224 |
| 15 | $(1, 1, 1, 1)$ | 23 | $(1, 1, 1, 1)$ | 232 |

# 6. PROPOSED MULTI-TERMINAL BAT

The proposed novel all-multiterminal BAT integrated with the TLSA and the proper-subset BAT is discussed in this section together with a discussion of its time complexity and an example to demonstrate its implementation.

## 6.1 Pseudocode and Time Complexity of Multi-terminal BAT

The procedure for the proposed all-multiterminal BAT can be divided into three main steps.

1. The traditional arc-based BAT [47] generates each binary-state vector, say $X$.

2. The proposed TLSA is implemented to find each connected node subset, for example, $X_j$ for $j = 0$, 1, 2, …, $(n_j - 1)$, in $G(X)$.

3. The proper-subset BAT is implemented to find each vectorized proper subset, say $Y_k$ for $k = 0$, 1, 2, …, $(n_k - 1)$, in $X_j$, and $\Pr(X)$ is added to $R_{|Y_k|}$.

The pseudocode of the proposed all-multiterminal BAT for calculating all multi-terminal reliabilities, that is, all-multiterminal reliability, in a binary-state network is presented as follows:

**Algorithm: All-Multiterminal BAT**

**Input:**      $G(V, E, \mathbf{D})$.

**Output:**    $R_\varsigma$ for each proper subset $\varsigma \subseteq V$.

**STEP 0.**    Let $i = R_\varsigma = 0$ for $\varsigma = 0$, 1, 2, …, $2^{(|V|-1)}$ and $X$ be a $m$-tuple vector zero.

**STEP 1.**    If $X(a_i) = 0$, let $X(a_i) = 1$ and go to STEP 3.

**STEP 2.**    Let $X(a_i) = 0$, $i = i + 1$, and go to STEP 1.

**STEP 3.**    If $i = m$, halt.

**STEP 4.**    Implement TLSA to find all connected nodes subsets: $X_j$ for $j = 0$, 1, 2, …, $(n_j - 1)$.

**STEP 5.**    Let $j = 0$.

**STEP 6.**    Implement the proposed proper-subset BAT to find a new vectorized node subset, say $Y_k$, in $X_j$. If no such $Y_k$, go to STEP 8.

**STEP 7.**    Calculate $|Y_k|$, let $R_{|Y_k|} = R_{|Y_k|} + \Pr(X)$ and go to STEP 6.

**STEP 8.**    If $j < n_j$, let $j = j + 1$ and go to STEP 6. Otherwise, go to STEP 1.

In the proposed all-multiterminal BAT, the loop from STEPs 0 to 3 finds each state vector $X$ based on the traditional arc-based BAT with time complexity $O(2^{m+1})$. STEP 4 finds each connected node proper subset $X_j$ based on the proposed TLSA with time complexity $O(n)$. The loop from STEPs 5 to 8 searches for all vectorized proper node subsets $Y_{j,k}$ in $X_j$ and add $\Pr(X)$ to $R_{|Y_{j,k}|}$ based on the proposed proper-subset BAT with time complexity $O(2^{n+1})$ [37]. Hence, the total time complexity is $O(n2^{n+1}2^{m+1})$ for the proposed all-multiterminal BAT in calculating the reliability of each node subset, that is, all-multiterminal, in a binary-state network.

### 6.2 Step-by-step Example

It is already NP-hard to calculate the multi-terminal reliability of a binary-state network, that is, the computational difficulty grows exponentially with the network size when calculating the multi-terminal reliability of a binary-state network [2]. Owing to the inherent NP-hard problem, to explain the general procedure of the proposed all-multiterminal BAT, the calculation of all-multiterminal reliability in a binary-state network is best illustrated with a simple example. Hence, instead of exhibiting practically large-scale networks, the benchmark network presented in Fig. 1 is chosen to establish the proposed methodology as in all existing algorithms for various network reliability problems [2, 3, 4].

The step-by-step procedure of the proposed all-multiterminal BAT is implemented to find all multiterminal reliabilities in Fig. 1.

**Solution:**

**STEP 0.**   Let $i = R_\varsigma = 0$ for $\varsigma = 1, 2, \ldots, 2^{(|V|-1)}$, and $X = (0, 0, 0, 0, 0)$ be a 5-tuple vector zero.

**STEP 1.**   Because $X(a_0) = 0$, let $X(a_0) = 1$, i.e., $X = (1, 0, 0, 0, 0)$, and go to STEP 3.

**STEP 3.**   Because $i = 0 < m = 5$, go to STEP 4.

**STEP 4.**   There is only one connected node subset: $X_0 = \{0, 1\}$ and $n^* = 1$.

**STEP 5.**   Let $j = 0$.

**STEP 6.**   $Y_k = (0, 1)$ after implementing the proposed proper-subset BAT.

**STEP 7.**   $|Y_k| = 0 + 2^1 = 2$, let $R_2 = R_2 + \Pr(X) = 0.9 \times 0.1^4 = 0.00009$ and go to STEP 8.

**STEP 8.** Because $j = (n^* - 1) = 0$, go to STEP 1.

**STEP 1.** Because $X(a_0) = 1$ in $X = (1, 0, 0, 0, 0)$, go to STEP 2.

**STEP 2.** Let $X(a_0) = 0$, $i = i + 1 = 1$, and go to STEP 1.

**STEP 1.** Because $X(a_1) = 0$, let $X(a_1) = 1$ and go to STEP 3, i.e., $X = (0, 1, 0, 0, 0)$.

**STEP 3.** Because $i = 1 < m = 5$, go to STEP 4.

**STEP 4.** There is only one connected node subset: $X_0 = \{0, 2\}$ and $n^* = 1$.

$$\vdots$$

$$\vdots$$

**STEP 1.** Because $X(a_0) = 0$, let $X(a_0) = 1$ and go to STEP 3, i.e., $X = (1, 0, 0, 0, 1)$.

**STEP 3.** Because $i = 0 < m = 5$, go to STEP 4.

**STEP 4.** After implementing the proposed TLSA, we have two connected nodes subsets: $X_0 = \{0, 1\}$, $X_1 = \{2, 3\}$, and $n^* = 2$.

**STEP 5.** Let $j = 0$.

**STEP 6.** We have $Y_0 = (0, 1)$ after implementing the proposed proper-subset BAT to $X_0 = \{0, 1\}$.

**STEP 7.** $|Y_0| = 1 + 2^1 = 3$, let $R_3 = R_3 + \Pr(X)$, where $\Pr(X) = 0.9^2 \times 0.1^3 = 0.00081$, and go to STEP 8.

**STEP 8.** Because $j = 0 < (n^* - 1) = 1$, let $j = j + 1 = 1$ and go to STEP 6.

**STEP 6.** $Y_1 = (2, 3)$ after implementing the proposed proper-subset BAT to $Y_1 = \{2, 3\}$.

**STEP 7.** $|Y_1| = 2^2 + 2^3 = 12$, let $R_{12} = R_{12} + \Pr(X)$, where $\Pr(X) = 0.9^2 \times 0.1^3 = 0.00081$, and go to STEP 8.

**STEP 8.** Because $j = (n^* - 1) = 1$, go to STEP 1.

$$\vdots$$
$$\vdots$$

The results of each multi-terminal reliability based on the proposed all-multiterminal BAT are listed in Table 7, and the corresponding node subsets of notations from column 4 entitled "3" to the last column entitled "15" are provided in Table 8, where $D_{avg}(V_i)$ is the average degree of node subset $V_i$. For example, the value in the cell of $i = 3$ and column 4 entitled "3" is 1, and it represents the node subset {0, 1} is connected in $X_i = X_3 = (1, 1, 0, 0, 0)$. In addition, $Y_3 = (1, 1, 0, 0)$ is the vectorized node subset corresponding to node subset {0, 1}, as shown in Table 8.

**Table 7.** Results for each multi-terminal reliability based on the proposed all-multiterminal BAT.

| $i$ | $X_i$ | $\mathrm{Pr}_{0.9}(X_i)$ | 3 | 5 | 6 | 7 | 9 | 10 | 11 | 12 | 13 | 14 | 15 |
|---|---|---|---|---|---|---|---|---|---|---|---|---|---|
| 0 | (0, 0, 0, 0, 0) | 0.00001 | | | | | | | | | | | |
| 1 | (1, 0, 0, 0, 0) | 0.00009 | | | | | | | | | | | |
| 2 | (0, 1, 0, 0, 0) | 0.00009 | | | | | | | | | | | |
| 3 | (1, 1, 0, 0, 0) | 0.00081 | 1 | 1 | 1 | 1 | | | | | | | |
| 4 | (0, 0, 1, 0, 0) | 0.00009 | | | | | | | | | | | |
| 5 | (1, 0, 1, 0, 0) | 0.00081 | 1 | 1 | 1 | 1 | | | | | | | |
| 6 | (0, 1, 1, 0, 0) | 0.00081 | 1 | 1 | 1 | 1 | | | | | | | |
| 7 | (1, 1, 1, 0, 0) | 0.00729 | 1 | 1 | 1 | 1 | | | | | | | |
| 8 | (0, 0, 0, 1, 0) | 0.00009 | | | | | | | | | | | |
| 9 | (1, 0, 0, 1, 0) | 0.00081 | 1 | | | | 1 | 1 | 1 | | | | |
| 10 | (0, 1, 0, 1, 0) | 0.00081 | | 1 | | | | 1 | | | | | |
| 11 | (1, 1, 0, 1, 0) | 0.00729 | 1 | 1 | 1 | 1 | 1 | 1 | 1 | 1 | 1 | 1 | 1 |
| 12 | (0, 0, 1, 1, 0) | 0.00081 | | | 1 | | | 1 | | 1 | | 1 | |
| 13 | (1, 0, 1, 1, 0) | 0.00729 | 1 | 1 | 1 | 1 | 1 | 1 | 1 | 1 | 1 | 1 | 1 |
| 14 | (0, 1, 1, 1, 0) | 0.00729 | 1 | 1 | 1 | 1 | 1 | 1 | 1 | 1 | 1 | 1 | 1 |
| 15 | (1, 1, 1, 1, 0) | 0.06561 | 1 | 1 | 1 | 1 | 1 | 1 | 1 | 1 | 1 | 1 | 1 |
| 16 | (0, 0, 0, 0, 1) | 0.00009 | | | | | | | | | | | |
| 17 | (1, 0, 0, 0, 1) | 0.00081 | 1 | | | | | | | 1 | | | |
| 18 | (0, 1, 0, 0, 1) | 0.00081 | | 1 | | | 1 | | | 1 | 1 | | |
| 19 | (1, 1, 0, 0, 1) | 0.00729 | 1 | 1 | 1 | 1 | 1 | 1 | 1 | 1 | 1 | 1 | 1 |
| 20 | (0, 0, 1, 0, 1) | 0.00081 | | | 1 | | | 1 | | 1 | | 1 | |
| 21 | (1, 0, 1, 0, 1) | 0.00729 | 1 | 1 | 1 | 1 | 1 | 1 | 1 | 1 | 1 | 1 | 1 |
| 22 | (0, 1, 1, 0, 1) | 0.00729 | 1 | 1 | 1 | 1 | 1 | 1 | 1 | 1 | 1 | 1 | 1 |
| 23 | (1, 1, 1, 0, 1) | 0.06561 | 1 | 1 | 1 | 1 | 1 | 1 | 1 | 1 | 1 | 1 | 1 |
| 24 | (0, 0, 0, 1, 1) | 0.00081 | | | 1 | | | 1 | | 1 | | 1 | |
| 25 | (1, 0, 0, 1, 1) | 0.00729 | 1 | 1 | 1 | 1 | 1 | 1 | 1 | 1 | 1 | 1 | 1 |
| 26 | (0, 1, 0, 1, 1) | 0.00729 | 1 | 1 | 1 | 1 | 1 | 1 | 1 | 1 | 1 | 1 | 1 |
| 27 | (1, 1, 0, 1, 1) | 0.06561 | 1 | 1 | 1 | 1 | 1 | 1 | 1 | 1 | 1 | 1 | 1 |
| 28 | (0, 0, 1, 1, 1) | 0.00729 | | | 1 | | | 1 | | 1 | | 1 | |
| 29 | (1, 0, 1, 1, 1) | 0.06561 | 1 | 1 | 1 | 1 | 1 | 1 | 1 | 1 | 1 | 1 | 1 |
| 30 | (0, 1, 1, 1, 1) | 0.06561 | 1 | 1 | 1 | 1 | 1 | 1 | 1 | 1 | 1 | 1 | 1 |
| 31 | (1, 1, 1, 1, 1) | 0.59049 | 1 | 1 | 1 | 1 | 1 | 1 | 1 | 1 | 1 | 1 | 1 |

**Table 8.** All node subsets in Fig. 1 obtained from the BAT.

| $i$ | $Y_i$ | $/Y_i/$ | $V_i$ | $|V_i|$ | $R_i = R(V_i)$ | $\mathrm{D}_{\mathrm{avg}}(V_i)$ |
|---|---|---|---|---|---|---|
| 0 | (0, 0, 0, 0) | 0 | $\varnothing$ | 0 | | 0 |
| 1 | (1, 0, 0, 0) | 1 | {0} | 1 | | 2/2 = 1 |
| 2 | (0, 1, 0, 0) | 2 | {1} | 1 | | 3/2 = 1.5 |
| 3 | (1, 1, 0, 0) | 3 | {0, 1} | 2 | 0.98820 | 4/2 = 2 |
| 4 | (0, 0, 1, 0) | 4 | {2} | 1 | | 3/2 = 1.5 |
| 5 | (1, 0, 1, 0) | 5 | {0, 2} | 2 | 0.98820 | 4/2 = 2 |
| 6 | (0, 1, 1, 0) | 6 | {1, 2} | 2 | 0.99630 | 3/2 = 1.5 |
| 7 | (1, 1, 1, 0) | 7 | {0, 1, 2} | 3 | 0.98658 | 5/4 = 1.25 |
| 8 | (0, 0, 0, 1) | 8 | {3} | 1 | | 2/2 = 1 |
| 9 | (1, 0, 0, 1) | 9 | {0, 3} | 2 | 0.97848 | 4/2 = 2 |
| 10 | (0, 1, 0, 1) | 10 | {1, 3} | 2 | 0.98820 | 4/2 = 2 |
| 11 | (1, 1, 0, 1) | 11 | {0, 1, 3} | 3 | 0.97767 | 5/3 = 1.66 |
| 12 | (0, 0, 1, 1) | 12 | {2, 3} | 2 | 0.98820 | 4/2 = 2 |
| 13 | (1, 0, 1, 1) | 13 | {0, 2, 3} | 3 | 0.97767 | 5/3 = 1.66 |
| 14 | (0, 1, 1, 1) | 14 | {1, 2, 3} | 3 | 0.98658 | 5/3 = 1.66 |
| 15 | (1, 1, 1, 1) | 15 | {0, 1, 2, 3} | 4 | 0.97686 | 5/4 = 1.25 |

From Table 7, we can find that

1. $R(V^*) = 0$ if $D_{avg}(V^*) = 0$, where $V^*$ is either an empty set or all nodes in $V^*$ are disconnected. Note that $R(V^*)$ is not necessarily zero if $D_{avg}(V^*) < 1$, that is, $D_{avg}(\{0, 1, 2, 3\}) = 3/4 < 1$, but $R(\{0,1,2,3\}) \neq 0$ in Fig. 5.

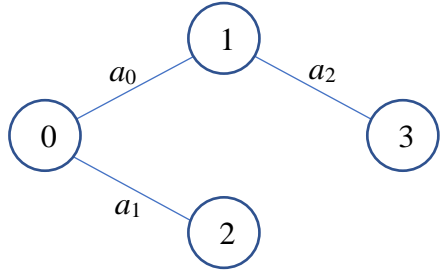

**Figure 5.** Example network to explain $D_{vag}(\{0, 1, 2, 3\}) < 1$ but $R_{\{0, 1, 2, 3\}} \neq 0$.

2. $R_0 = R_1 = R_2 = R_4 = R_8 = 0$ because $R(\varnothing) = R(\{0\}) = R(\{1\}) = R(\{2\}) = R(\{3\}) = 0$, that is, a multi-terminal is a node subset with at least two nodes.

3. Multi-terminals with a larger average degree tend to have a higher chance of obtaining better reliabilities than those with a smaller average degree. For example, in Fig. 1 and Table 7, nodes 1 and 2 both have the highest degree 3, that is, the multi-terminal $\{1, 2\}$ with the highest average degree of 2.5, and $R(\{1, 2\}) = R_6 = 0.99630$ is better than the others.

4. In general, the larger the number of nodes in a multi-terminal, the smaller the reliability, and vice versa. For example, $\frac{\sum_{U_2} R(U_2)}{(\text{the number of } U_2)} = 0.98793 > \frac{\sum_{U_3} R(U_3)}{(\text{the number of } U_3)} = 0.982125 > \frac{\sum_{U_4} R(U_4)}{(\text{the number of } U_4)} = 0.976860$ in Fig. 1, where $U_k$ is all node subsets with $k$ nodes, for example, $U_2 = \{0, 1\}, \{0, 2\}, \{0, 3\}, \{1, 2\}, \{1, 3\},$ and $\{2, 3\}$.

5. In a symmetry graph, if $V^*$ is symmetric to $V^\#$, $R(V^*) = R(V^*)$. For example, Fig. 1 is a horizontal symmetric graph and a vertical symmetric graph that can be observed after drawing a horizontal line (see Fig. 6(a)) or a vertical line (see Fig. 6(b)), respectively. Hence, $R(\{0, 1\}) = R(\{2, 3\}) = R(\{0, 1\}) = R(\{0, 2\}) = R(\{1, 3\}) = 0.98820, R(\{0, 1, 2\}) = R(\{1, 2, 3\}) = 0.98658,$ and $R(\{0, 1, 3\}) = R(\{0, 2, 3\}) = 0.97767.$

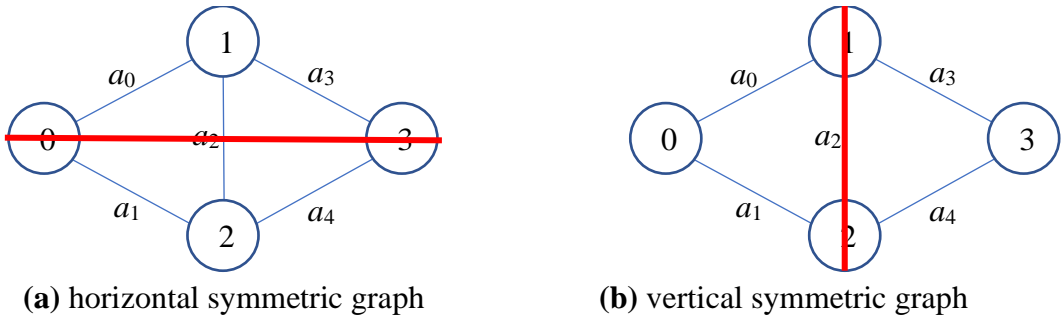

**(a)** horizontal symmetric graph　　　**(b)** vertical symmetric graph

**Figure 6.** horizontal and vertical symmetric graphs.

Note that:

1. Point 4 is critical and shows that the two-terminal reliability is higher than that of the multi-terminal reliability with more than two nodes. However, in real-life applications, such as smart grids [9], supply chains [4], and the Internet of Things [11, 12], multi-terminal reliability problems are more pragmatic and suitable. Hence, this study focuses on all-multiterminal reliability problems that are very valuable, visionary, and practical.

2. Point 5 is very useful in real-world applications [53], for example, instance cluster synchronization [54], network structural [55], and network spectral [56], because it has been revealed that real-life networks possess numerous symmetries [56].

## 7. CONCLUSIONS

Reliability is generally chosen as one of the most significant evaluation indices for real-world network applications. The major methods for evaluating network reliability are focused on the connectivity of two terminals, that is, two-terminal reliability. To meet the practical needs of real-world complex networks, the proposed all-multiterminal reliability problem generalizes the two-terminal reliability to multi-terminal reliability to consider the reliability among a proper node subset in binary-state networks, and then further extended to all-multiterminal reliability to study all multi-terminal reliabilities.

All-multiterminal reliability has practical effects that have not yet been explored in full generality, nor exploited by network practitioners. To the author's knowledge, the proposed all-multiterminal BAT is the first BAT and the first algorithm to find each multi-terminal reliability of binary-state networks simultaneously.

The proposed all-multiterminal BAT integrates the arc-based BAT in finding each state vector; the proposed TLSA finds all connected disjoint node subsets by way of exploiting the structure of the network constructed from the state vector; the proposed proper-subset BAT used for finding each proper node subset of connected node subsets and calculate the corresponding reliability of each proper node subset.

The proposed all-multiterminal BAT proved to be useful in terms of time complexity. The results and observations from this example are noteworthy and widely applicable. Therefore, we are confident that in the future, the proposed algorithm can be applied to various real-life applications.

**ACKNOWLEDGMENT**

This research was supported in part by the Ministry of Science and Technology, R.O.C. under grant MOST 107-2221-E-007-072-MY3 and MOST 110-2221-E-007-107-MY3. This article was once submitted to arXiv as a temporary submission that was just for reference and did not provide the copyright.